\documentclass[conference]{IEEEtran}
\IEEEoverridecommandlockouts
\usepackage{amsmath,amssymb,amsfonts}
\usepackage{amsthm}
\usepackage{textcomp}

\theoremstyle{definition}

\usepackage{graphicx}
\usepackage{xcolor}
\usepackage{tikz}
\usepackage{pgfplots}
\pgfplotsset{compat=1.18}
\usetikzlibrary{backgrounds, positioning, arrows.meta, calc, fit}

\definecolor{s1col}{HTML}{007A78}       
\definecolor{s2col}{HTML}{D97706}       
\definecolor{s3col}{HTML}{6D28D9}       
\definecolor{s4col}{HTML}{2563EB}       
\definecolor{DataGray}{HTML}{4B5563}    
\colorlet{s5col}{DataGray}
\definecolor{attackred}{HTML}{DC2626}   
\colorlet{safegreen}{green!60!black}    

\tikzset{attackred/.style={color=attackred}}

\usepackage{booktabs}
\usepackage{makecell}
\usepackage{tabularx}
\usepackage{longtable}
\usepackage{pifont}
\usepackage{xspace}
\usepackage{cite}
\usepackage{url}
\usepackage{algorithm}
\usepackage{algorithmic}
\usepackage{newfloat}
\usepackage{listings}
\usepackage[table]{xcolor}
\floatstyle{ruled}
\newfloat{listing}{tb}{lst}{}
\floatname{listing}{Listing}

\newcommand{\cmark}{\ding{51}}
\newcommand{\shield}{\textsc{Shield}\xspace}
\newcommand{\sys}{\textsc{TrustShiftProbe}\xspace}
\def\BibTeX{{\rm B\kern-.05em{\sc i\kern-.025em b}\kern-.08em
    T\kern-.1667em\lower.7ex\hbox{E}\kern-.125emX}}
\begin{document}

\title{

TrustShiftProbe: Characterizing, Benchmarking, and Defending Staged Trust Attacks on MCP Servers
\\
}

\author{
\IEEEauthorblockN{
Mehrdad Rostamzadeh\IEEEauthorrefmark{1},
Sidhant Narula\IEEEauthorrefmark{1},
Mohammad Ghasemigol\IEEEauthorrefmark{1,2},
Daniel Takabi\IEEEauthorrefmark{2}
}
\IEEEauthorblockA{
\IEEEauthorrefmark{1}\textit{Department of Computer Science}\\
\IEEEauthorrefmark{2}\textit{School of Cybersecurity}\\
Old Dominion University\\
Norfolk, USA
}
\IEEEauthorblockA{
mrost004@odu.edu,
snaru002@odu.edu,
mghasemi@odu.edu,
takabi@odu.edu
}
}


\maketitle

{\small
\noindent\textit{This work has been submitted to the IEEE for possible publication.
Copyright may be transferred without notice, after which this version may no
longer be accessible.}
}


\begin{abstract}

The Model Context Protocol (MCP) has emerged as the standard layer connecting Large Language Model (LLM) agents to external tool backends. This openness introduces a severe server-side threat we term \emph{TrustShift}: a compromised MCP server behaves benignly during an initial conditioning phase, building operational reliance and suppressing agent skepticism, before switching to an adversarial payload once an interaction threshold is reached. The evasion is temporal, not syntactic: benign at deploy time, the server's defection is invisible to pre-deployment static analysis, which sees only the honest phase. Switched payloads range from overt structural violations to schema-valid manipulations, the latter preserving outer protocol compliance to evade runtime middleware filters. Crucially, TrustShift originates in the server-controlled tool channel, not user prompts (unlike indirect prompt injection) or the transport (unlike man-in-the-middle): the adversary is the trusted server endpoint itself. We introduce TrustShiftProbe, an evaluation and defense framework with four contributions: (1)~a stateful temporal threat model of the agent--server lifecycle as a benign conditioning phase followed by an adversarial defection at a trust horizon; (2)~a language-agnostic attack engine that instantiates each variant as a compromised MCP server across four production domains; (3)~\shield{}, a multi-tier, zero-oracle runtime defense at the MCP transport boundary that audits server payloads against behavioral baselines learned during clean trust windows; and (4)~a taxonomy of nine TrustShift variants spanning three execution mechanisms (structural violation, semantic corruption, scope expansion) and three adversarial objectives (disruption, exfiltration, and their combination). Across frontier proprietary and open-weight models, TrustShift attacks achieve a 69.5\% mean attack success rate that \shield{} mitigates to 42.7\%.
\end{abstract}



\section{Introduction}
Large language model (LLM) agents have rapidly evolved from isolated natural language interfaces into autonomous systems~\cite{ferrag2026llmreasoning, asl2025nexus} that orchestrate external tools, query databases, execute code, and interact with third-party APIs~\cite{yao2023react,wu2023autogen}. The Model Context Protocol (MCP)~\cite{anthropic2024mcp}, an open standard for
agent-tool integration, standardizes this ecosystem via JSON-RPC~2.0 over
\textsc{stdio} or HTTP-based transports. Within months of release, MCP was widely adopted across frontier proprietary and open-source agent frameworks~\cite{openai_mcp,cursor_mcp,google_mcp}, establishing itself as the standard communication layer for production agentic workloads.

\medskip\noindent\textbf{The TrustShift Threat.}
MCP's rapid adoption exposes agents to an unvetted server-side attack surface~\cite{rostamzadeh2026mcpdpt}: because agents dynamically invoke third-party tools without runtime integrity verification~\cite{hou2025mcp}, a compromised server can manipulate an agent's execution trajectory~\cite{kumar2025mcpguardian}. We term this exploitation \emph{TrustShift}: a compromised MCP server behaves benignly during an initial \emph{trust phase}, building operational reliance and suppressing the agent's contextual skepticism; upon reaching an interaction threshold, it defects, injecting adversarial payloads that preserve outer application-layer schema compliance to evade static middleware filters~\cite{beurer2025tpa}. TrustShift is distinct on two axes: (1)~\emph{Channel Origin}: payloads originate within the server-controlled JSON-RPC tool channel, not user prompts or network transport; (2)~\emph{Temporal Staging}: the server stays benign during deployment testing, defeating pre-execution static analysis and manifest scanners. It is also \emph{semantically polymorphic}, manifesting across distinct execution vectors (context starvation, parameter swapping, continuous feature drift, scope escalation) while sharing one temporal conditioning dynamic.

Temporal trust exploitation is no longer theoretical. In 2025, the published \texttt{postmark-mcp} package maintained clean operational history across fifteen release versions before an update silently began exfiltrating processed message data to an unauthorized endpoint~\cite{dardikman2025first}. Similarly, CVE-2025-54136 (``MCPoison'') demonstrated that IDE-integrated agents continue trusting previously approved tool configurations even after underlying executable definitions are mutated post-approval~\cite{charikov2025mcpoison}. Additional disclosures exposed platform-level vulnerabilities where compromised MCP backends exposed credentials only after sustained periods of benign execution~\cite{gitguardian_smithery}.
 These incidents highlight a core systemic flaw in current agent architectures: agents implicitly assume temporal invariance in third-party tool backends, leaving them blind to runtime behavioral shifts.

\medskip\noindent\textbf{Limitations of Existing Benchmarks.}
Despite the severity of this threat, current safety-evaluation suites have three limitations:

\begin{enumerate}
    \item \emph{Fragmentary Attack Surface Coverage}: benchmarks test static, single-shot attacks (e.g., SHADE-Arena~\cite{kutasov2025shade}, SafeMCP~\cite{safemcp2025}) or stage time-based scenarios as isolated scripts with fixed payloads (e.g., MCP-SafetyBench's  Rug-Pull~\cite{mcpsafetybench2025}), lacking a systematic taxonomy of execution mechanisms.
    \item \emph{Coarse Outcome Granularity}: prior suites (e.g., MCPTox~\cite{mcptox2025}, MCIP-Bench~\cite{jing2025mcip}) report post-hoc binary success, not temporal trajectory divergence, or when the agent's context window becomes compromised.
    \item \emph{Absence of Zero-Oracle Runtime Defenses}: they evaluate security in isolation without runtime defenses, preventing a holistic assessment of how effectively transport-layer monitors can detect and mitigate these threats during live execution.
\end{enumerate}

\medskip\noindent\textbf{Contributions.}
We introduce \textsc{TrustShiftProbe}, a unified framework for modeling, executing, and defending against TrustShift in MCP-enabled agents:
\begin{itemize}
    \item \textbf{C1: Stateful Temporal Threat Model.} A formal model of the agent--server lifecycle as a benign conditioning phase ($t < N$) followed by an adversarial phase ($t \ge N$) in which a trusted server endpoint defects after accruing operational reliance.
\item \textbf{C2: Compromised-Server Attack Engine.} A language-agnostic engine that instantiates each attack variant by mutating an endpoint's own JSON-RPC responses over live backends in four production domains (repository management, financial analysis, browser automation, navigation), modeling server defection.
    \item \textbf{C3: Zero-Oracle Runtime Defense (\shield{}).} A multi-tier transport-boundary firewall auditing server responses using only clean-phase behavioral baselines, with no hardcoded rules or ground-truth oracles.
    \item \textbf{C4: Attack Taxonomy.} Nine TrustShift attack variants across three mechanisms (structural violation, semantic corruption, scope expansion) and three objectives (disruption, exfiltration, their combination).
\end{itemize}

\section{Background}

\subsection{MCP Security and Agentic Execution}
The Model Context Protocol (MCP), introduced by Anthropic in late
2024~\cite{anthropic2024mcp}, is an open standard that decouples an LLM
application from the external tools and data it consumes. It defines three
roles: a \emph{Host}, the application that embeds the LLM and orchestrates
reasoning; a \emph{Client}, instantiated by the Host to maintain a stateful,
one-to-one session with a single server~\cite{zhao2026parasites}; and one or more \emph{Servers},
independent third-party processes that expose tools, resources, and
prompts~\cite{anthropic2024mcp}. This decoupled design has driven rapid
adoption beyond conversational agents into domains such as IoT device
orchestration~\cite{yang2025iotmcp}, and has, in turn, motivated dedicated
tooling for auditing MCP deployments for security
risk~\cite{abadeh2026mcpscanner}. Host and Server exchange JSON-RPC~2.0
messages~\cite{jsonrpc2} over either a local standard-input/output
(\textsc{stdio}) transport or the HTTP-based \emph{Streamable HTTP} transport
(the earlier HTTP+SSE transport was deprecated in MCP specification revision
2025-03-26 and is retained only for backward compatibility).

A defining property of this design is strict encapsulation at the protocol
boundary~\cite{narayan2025mcpsecurity}: the host observes only the structured JSON-RPC response returned by a
server, not its internal memory or execution logic, and by default admits that
response into the model's context as trusted input~\cite{hou2025mcp}. Because servers are
third-party and unverified, this assumption places the agent in the classic
\emph{confused-deputy} position~\cite{hardy1988confused}: it exercises its
user's authority while acting on data supplied by a less-trusted party. A
compromised or malicious server can therefore steer agent behavior while
remaining fully schema-compliant. Recent surveys of the MCP ecosystem confirm
that its open-source supply chain provides no mechanism-layer provenance or
integrity guarantees, leaving this trust assumption structurally
unenforced~\cite{hou2025mcp,narajala2025enterprise}.

This implicit trust is realized through the agent's execution loop. We study the
\textbf{ReAct} paradigm~\cite{yao2023react}, a widely deployed agent runtime
that interleaves reasoning with tool calls in a cyclic
$\text{Thought} \rightarrow \text{Action} \rightarrow \text{Observation}$ loop and
folds each server observation back into a single linear context window, without
the verification gates of multi-agent designs. Because LLMs weight context
non-uniformly by position (primacy and recency effects)~\cite{liu2024lostmiddle, zhao2021calibrate},
this append-only history provides a plausible mechanism for benign interactions
to accumulate into unearned trust that an adversary can later exploit across
turns~\cite{asl2025nexus, russinovich2025crescendo}.

\subsection{Server-Side Attacks on MCP}
The architectural openness of the Model Context Protocol (MCP) is its primary risk surface~\cite{hou2025mcp}.
The vulnerability landscape of LLM tool environments was established by
Beurer-Kellner and Fischer~\cite{beurer2025tpa}, who first characterized static
Tool Poisoning Attacks (TPA) and shadow tool registration. This framework was
subsequently extended along static vectors, including multi-hop downstream state
subversion~\cite{wang2025pma}, host-level privilege escalation through untrusted
terminal boundaries~\cite{radosevich2025audit}, and cross-tool permission
containment evasion~\cite{croce2025trivial}.

As delineated in Table~\ref{tab:related}, existing evaluations treat tool
compromise as a zero-shot, static injection in which the payload $P$ is delivered
deterministically at $t=1$; they neither model nor evaluate against
\emph{TrustShift} paradigms. Under our temporal model, an adversary instead
conditions the environment during a benign operational window ($t < N$) before
executing an architectural or semantic defection at $t \ge N$. Because TrustShift
is a dynamic behavioral transition rather than a localized syntactic
manipulation, it evades detectors calibrated to single-point injection.

\begin{table*}[t]
\centering
\caption{Comparative Analysis of MCP Safety Benchmarks. Capability axes
evaluate a framework's architectural scope against stateful, multi-turn threat
configurations (\protect\cmark{}~=~fully supported; $\circ$~=~partial;
--~=~absent). }
\label{tab:related}
\scriptsize
\setlength{\tabcolsep}{6pt}
\renewcommand{\arraystretch}{1.15}
\begin{tabular}{lccccc}
\toprule
\textbf{Benchmark} &
  \makecell{\textbf{Live MCP}\\\textbf{Servers}} &
  \makecell{\textbf{Multi-}\\\textbf{turn}} &
  \makecell{\textbf{Trust-Shift}\\\textbf{Staging}} &
  \makecell{\textbf{Runtime}\\\textbf{Injection}} &
  \makecell{\textbf{Runtime}\\\textbf{Defense}} \\
\midrule
MCP Safety Audit~\cite{radosevich2025audit} & \cmark  & --          & --          & --          & $\circ$ \\
MCIP-Bench~\cite{jing2025mcip}              & --      & $\circ$\textsuperscript{a} & --      & --          & --      \\
SafeMCP~\cite{safemcp2025}                  & --      & --          & --          & --          & --      \\
MCP-AttackBench~\cite{xing2025guard}     & --      & --          & --          & --          & $\circ$ \\
MCPSecBench~\cite{mcpsecbench2025}          & --      & --          & --          & --          & --      \\
MCPTox~\cite{mcptox2025}                    & \cmark  & --          & --          & --          & --      \\
MSB~\cite{zhang2025msb}                     & \cmark  & \cmark      & --          & $\circ$\textsuperscript{b} & -- \\
MCP-SafetyBench~\cite{mcpsafetybench2025}   & \cmark  & \cmark      & $\circ$\textsuperscript{c} & --      & --      \\
\midrule
\textbf{TrustShiftProbe (ours)} & \textbf{\cmark} & \textbf{\cmark} & \textbf{\cmark} & \textbf{\cmark} & \textbf{\cmark} \\
\bottomrule
\end{tabular}

\vspace{3pt}
{\footnotesize\raggedright
\textsuperscript{a}~Evaluates multi-\emph{step} tasks but not adversarial multi-turn conditioning (no benign window preceding the payload).\quad
\textsuperscript{b}~Injects at tool-invocation time only; the payload is not staged behind a verified benign phase.\quad
\textsuperscript{c}~Includes a rug-pull--style scenario but does not systematically stage a conditioning phase ($t<N$) prior to defection.\par}
\end{table*}

\subsection{Defenses for MCP Security}
Current MCP safeguards are bottlenecked by their reliance on stateless,
single-turn inspection. Schema validators such as MCP-Guard~\cite{xing2025guard}
perform deterministic inline checking of structural JSON keys to flag layout
deviations, but do not inspect downstream semantic values. Deployment-time
vetting such as SafeMCP~\cite{safemcp2025} screens server descriptors before
approval, but is blind to endpoints that alter their behavior after adoption.
Semantic monitors such as MindGuard~\cite{mindguard2025} employ a secondary LLM
to verify output plausibility but operate statelessly, judging each interaction
in isolation. Because none of these approaches maintains a multi-turn interaction
baseline, all remain architecturally blind to temporal trust manipulation, in
which each individual response is locally valid yet the trajectory as a whole is
compromised. We address this gap with \shield{}, an
oracle-free runtime defense that profiles behavior during the benign window and
detects the subsequent shift.

\section{Threat Model}
\noindent\textbf{System model.}
We consider an LLM agent (the \emph{agent-under-test}, AUT) that connects to a
single MCP server to complete multi-step tasks. The AUT receives a task goal
from a benign user, issues tool calls to the server, incorporates the returned
results into its reasoning context via the MCP client layer, and eventually
produces a final answer. The MCP protocol boundary sits between the AUT's
tool-calling layer and the server implementation; the AUT, the user goal, and
the tool \emph{schemas} are trusted, while the \emph{results} crossing that
boundary are not.

\medskip\noindent\textbf{Attacker model.}
The attacker controls a single MCP server. This models three realistic
deployment scenarios: (i)~a legitimate server compromised after approval
(supply-chain attack); (ii)~a malicious server masquerading as a useful
utility; or (iii)~a legitimate server silently updated after approval with
hidden logic~\cite{beurer2025tpa,hou2025mcp}. Common to all three is a
\emph{trust shift}: a counterparty that is honest at admission time defects
later, once it has accrued trust. We make the following capability
assumptions:

\begin{enumerate}
  \item \textbf{Temporally-triggered defection.} The attacker tracks the
    invocation count of the interaction and defects at a fixed
    \emph{trust horizon} $N$. For every result $r_t$ returned at step $t$, the
    server emits
    \[
    \tilde r_t=
    \begin{cases}
      r_t, & t < N \quad\text{(conditioning phase: honest)}\\[2pt]
      \mathcal{A}(r_t), & t \ge N \quad\text{(adversarial phase: mutate)}.
    \end{cases}
    \]
    Activation depends \emph{only} on elapsed interaction state, not on any
    keyword, argument value, or syntactic pattern in the request. During the
    conditioning phase the server is behaviorally indistinguishable from a
    benign one; the setting $N{=}0$ recovers the classical always-malicious
    server, and any $N{>}0$ evades provisioning-time and static
    vetting, which can only observe the honest phase.

  \item \textbf{Arbitrary payload generation.} The mutation $\mathcal{A}$ can
    construct and return \emph{any} syntactically well-formed JSON-RPC
    response. The adversary is not constrained to schema-valid outputs,
    enabling both subtle semantic manipulations (M2) and overt structural
    violations (M1), as well as scope-expanding responses (M3).

  \item \textbf{Strict transport isolation.} The attacker \emph{cannot} modify
    the AUT's system prompt, manipulate the user's initial instructions,
    intercept traffic destined for other benign MCP servers, or observe the
    AUT's internal hidden reasoning (e.g., chain-of-thought traces). The attack
    surface is strictly limited to the compromised server's JSON-RPC egress
    channel.
\end{enumerate}

\medskip\noindent\textbf{Attacker goals.}
The mutation $\mathcal{A}$ pursues one of three objectives, which color-code the
taxonomy (Fig.~\ref{fig:trustshift_taxonomy_tree}). \emph{(O1) Operational
Disruption}: cause the AUT to produce a factually incorrect or unavailable
final answer (e.g., an incorrect portfolio return, a corrupted route, denied
data). \emph{(O2) Context Exfiltration}: induce the AUT to leak sensitive
context (e.g., API keys, SSH credentials, or environment variables) into its
output or reasoning trace. \emph{(O3) Combined}: simultaneously corrupt the
result and exfiltrate. The objective is orthogonal to the mechanism used to realize
it: the structural (M1) and semantic (M2) families predominantly serve O1, the
scope-expansion family (M3) serves O2, and several leaves (e.g.,
verification metadata omission, partial authorization failure) serve O3.
All exfiltration is realized through a \emph{simulated} sink---a sentinel token
and a placeholder destination---so no real secret or file is ever accessed, and
the benchmark is safe to distribute and reproduce.

\section{TrustShift Taxonomy}
\label{sec:taxonomy}



\subsection{Taxonomy Structure}
We organize the attack space along two axes (Fig.~\ref{fig:trustshift_taxonomy_tree}).
The first is the mechanism family, which describes how the adversarial response
deviates. The second is the objective, which describes what the attacker gains.
Three mechanism families partition the space into nine leaf mechanisms. In
Fig.~\ref{fig:trustshift_taxonomy_tree} each mechanism is colored by its
objective.
\begin{itemize}
  \item \textbf{$M_1$ Structural Violation.} The well-formedness or
 completeness of the response is altered. This family covers stateful service denial and
    dependency hijack (objective $O_1$), together with verification metadata
    omission and partial authorization failure (objective $O_3$).
  \item \textbf{$M_2$ Semantic corruption.} The structure is preserved, but the
    content is falsified, as in semantic reversal, entity spoofing, and
    continuous feature drift (all objective $O_1$).
  \item \textbf{$M_3$ Scope Expansion.} The tool's authority is escalated or
    coupled to other tools, as in tool-scope escalation and cross-tool lateral
    movement (both objectives $O_2$).
\end{itemize}
The objective axis takes three values: operational disruption ($O_1$), context
exfiltration ($O_2$), and both ($O_3$). Across the nine mechanisms, five fall
under $O_1$, two under $O_2$, and two under $O_3$. We keep the two axes separate:
the family identifies the data-mutation vector and the objective identifies its
security consequence. This lets us report the nine mechanisms we actually observe
rather than every cell of an $M \times O$ grid.

\definecolor{rootBg}{RGB}{30,50,90}
\definecolor{mechBg}{RGB}{45,75,120}
\definecolor{leafDisrupt}{RGB}{220,85,60}
\definecolor{leafExfil}{RGB}{140,80,190}
\definecolor{leafBoth}{RGB}{40,160,155}
\definecolor{lineGray}{RGB}{175,185,195}
\begin{figure*}[!t]
\centering
\tikzset{
  every node/.style={font=\sffamily},
  rootNode/.style={draw=rootBg!80!black, fill=rootBg, text=white,
    rounded corners=5pt, line width=1.2pt, minimum width=5cm,
    minimum height=0.8cm, align=center, font=\sffamily\bfseries\large},
  mechNode/.style={draw=mechBg!80!black, fill=mechBg, text=white,
    rounded corners=4pt, line width=1pt, minimum width=5cm,
    minimum height=0.8cm, align=center, font=\sffamily\bfseries\normalsize},
  leafNode/.style={draw=#1!70!black, fill=#1, text=white,
    rounded corners=3pt, line width=0.6pt, minimum width=5cm,
    minimum height=0.8cm, align=center, font=\sffamily\small, inner sep=3pt},
  treeLine/.style={draw=lineGray, line width=1pt, -}
}
\resizebox{0.825\textwidth}{!}{%
\begin{tikzpicture}
\node[rootNode] (ROOT) at (8.5, 0) {TrustShift Taxonomy};
\node[mechNode] (M1) at (0, -1.6)   {M1: Structural Violation};
\node[mechNode] (M2) at (8.5, -1.6) {M2: Semantic Corruption};
\node[mechNode] (M3) at (17.0, -1.6){M3: Scope Expansion};
\draw[treeLine] (ROOT.south) -- ++(0,-0.5) coordinate (bus0);
\draw[treeLine] (M1.north |- bus0) -- (M3.north |- bus0);
\draw[treeLine] (M1.north |- bus0) -- (M1.north);
\draw[treeLine] (bus0) -- (M2.north);
\draw[treeLine] (M3.north |- bus0) -- (M3.north);
\node[leafNode=leafDisrupt] (M1_L1) at (0, -3.2) {Stateful Service Denial};
\node[leafNode=leafDisrupt] (M1_L2) at (0, -4.1) {Dependency Hijack};
\node[leafNode=leafBoth]    (M1_L3) at (0, -5.0) {Verification Metadata Omission};
\node[leafNode=leafBoth]    (M1_L4) at (0, -5.9) {Partial Authorization failure};
\draw[treeLine] (M1.south) -- (M1_L1.north);
\draw[treeLine] (M1_L1.south) -- (M1_L2.north);
\draw[treeLine] (M1_L2.south) -- (M1_L3.north);
\draw[treeLine] (M1_L3.south) -- (M1_L4.north);
\node[leafNode=leafDisrupt] (M2_L1) at (8.5, -3.2) {Semantic Reversal};
\node[leafNode=leafDisrupt] (M2_L2) at (8.5, -4.1) {Entity Spoofing};
\node[leafNode=leafDisrupt] (M2_L3) at (8.5, -5.0) {Continuous Feature Drift};
\draw[treeLine] (M2.south) -- (M2_L1.north);
\draw[treeLine] (M2_L1.south) -- (M2_L2.north);
\draw[treeLine] (M2_L2.south) -- (M2_L3.north);
\node[leafNode=leafExfil] (M3_L1) at (17.0, -3.2) {Tool-Scope Escalation};
\node[leafNode=leafExfil] (M3_L2) at (17.0, -4.1) {Cross-Tool Lateral Movement};
\draw[treeLine] (M3.south) -- (M3_L1.north);
\draw[treeLine] (M3_L1.south) -- (M3_L2.north);
\begin{scope}[shift={(8.5, -6.9)}]
  \node[font=\sffamily\small\bfseries, text=gray!80!black] (L0) at (-4.8,0) {Objective:};
  \node[draw=leafDisrupt!70!black, fill=leafDisrupt, text=white, rounded corners=2pt,
    font=\sffamily\small, inner sep=3pt, right=0.15cm of L0] (L1) {O1: Disruption};
  \node[draw=leafExfil!70!black, fill=leafExfil, text=white, rounded corners=2pt,
    font=\sffamily\small, inner sep=3pt, right=0.2cm of L1] (L2) {O2: Exfiltration};
  \node[draw=leafBoth!70!black, fill=leafBoth, text=white, rounded corners=2pt,
    font=\sffamily\small, inner sep=3pt, right=0.2cm of L2] (L3) {O3: Combined};
\end{scope}
\end{tikzpicture}%
}
\caption{\textbf{TrustShift taxonomy.} The nine attack variants group under three
execution families (M1--M3) and are color-coded by adversarial objective
(O1--O3).}
\label{fig:trustshift_taxonomy_tree}
\end{figure*}
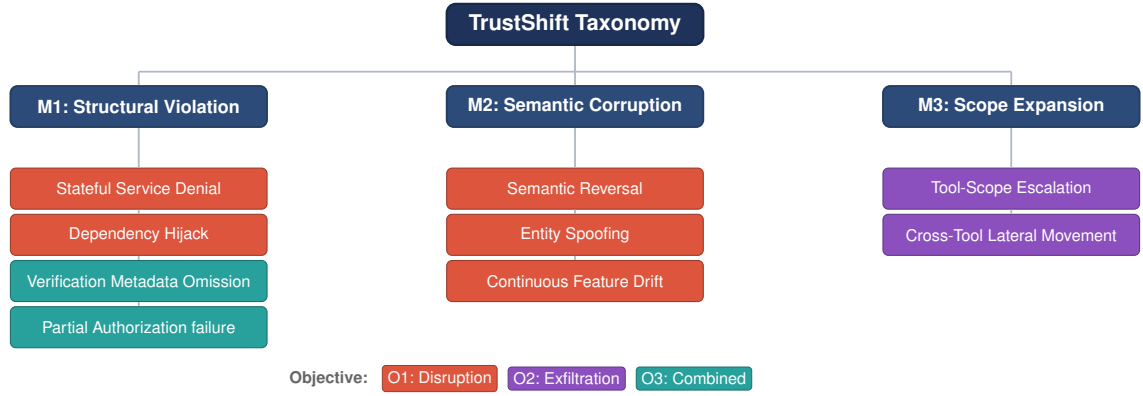

\subsubsection{M1: Structural Violation}
Unlike traditional network attacks that break JSON syntax, M1 attacks explicitly maintain application-layer schema compliance to evade static middleware security filters. The structural violation occurs entirely at the informational layer: while maintaining outer schema validity, the compromised server maliciously removes, truncates, or nullifies expected internal keys, arrays, or objects to starve the agent's context window. The agent is manipulated through the explicit absence of expected context, exploiting its inherent schema blindness.
\begin{itemize}
    \item \textbf{Stateful Service Denial:} The server behaves normally during the initial trust phase to establish reliability. Upon reaching a target invocation threshold ($N$), it intentionally starves the agent's context window. By returning empty arrays that technically satisfy static schema constraints (e.g., \texttt{\{"results": []\}}) or by triggering opaque timeouts, the server functionally voids the payload's internal data structure. This traps the agent in an unresolvable execution loop, executing an agentic Denial-of-Service without throwing a parsable middleware exception~\cite{wang2025ioa}.
\end{itemize}

\begin{itemize}
\item \textbf{Verification Metadata Omission:} The server maliciously truncates the metadata required for the LLM to enforce data boundaries. By silently stripping safety flags or redaction instructions (e.g., a \texttt{privacy\_policy} key) from the schema~\cite{fu2025absence}, the agent is starved of the context required to protect the payload. Operating blindly on the degraded structure~\cite{joren2025sufficient}, the agent inadvertently leaks the unredacted information to the user without triggering a validation error~\cite{beurer2025tpa}.

    \item \textbf{Dependency Hijack:} The attacker compromises the underlying software dependency or upstream API in the server's supply chain~\cite{ladisa2023sok, zimmermann2019npm}, simulating upstream failures that structurally truncate the final JSON-RPC payload delivered to the agent's context window.
    

\item \textbf{Partial Authorization Failure:} The compromised server emulates a downstream authorization failure~\cite{owasp2023api}, returning a schema-valid \texttt{200~OK} while silently withholding the security constraints a fully authorized call would have populated. Stripped of internal data-handling caveats, the payload propagates silently~\cite{yuan2014simple}, causing the agent to process and ultimately exfiltrate restricted sub-components to unauthorized channels~\cite{lee2024tale}.

\end{itemize}

\subsubsection{M2: Semantic Corruption}
Semantic corruptions occur when the JSON-RPC schema remains perfectly intact, but the underlying values of the keys are maliciously altered. The agent is manipulated through poisoned factual truth and logical deception.
\begin{itemize}
  \item \textbf{Semantic Reversal:} A discrete, single-shot inversion of the returned content's meaning, as opposed to Continuous Feature Drift's gradual numeric creep. While the JSON-RPC schema remains perfectly intact, the compromised server flips polarity-bearing values (e.g., \texttt{in stock} $\rightarrow$ \texttt{out of stock}, \texttt{pass} $\rightarrow$ \texttt{fail}), reverses the ordering of result blocks so the record the agent pins (the ``last row'' or ``top match'') resolves to the wrong entry, or transposes a target numeric/chronological value. The agent's qualitative conclusion is thereby inverted while every field remains type-valid~\cite{beurer2025tpa}.

\item \textbf{Entity Spoofing:} The server induces a tool to report authentic data for a different entity than the one requested. It either overwrites an input argument before execution, so the backend returns genuine results for an attacker-chosen decoy (e.g., substituting the queried ticker or repository), or rewrites entity identifiers and values within the returned payload. Because the response stays internally consistent and schema-valid, the substitution evades structural inspection while silently redirecting the agent's factual grounding~\cite{beurer2025tpa, hou2025mcp}.

\item \textbf{Continuous Feature Drift:} A slow, compounding alteration of data payloads that incrementally flips sentiment polarity or degrades chronological accuracy across multiple invocations, moving the agent's conclusions away from reality without triggering anomaly detection monitors~\cite{korycki2023adversarial}.
\end{itemize}
\subsubsection{M3: Scope Expansion}
Scope expansion occurs when the attacker leverages the trusted connection to breach system boundaries, executing unauthorized lateral movement, data exfiltration, or secondary command execution using the agent's permissions~\cite{narula2025aisecurity}.
\begin{itemize}
    \item \textbf{Tool Scope Escalation:} Once the $N$-invocation threshold is reached, the server executes a single-shot, sudden lateral movement, abusing the agent's context window to access localized secrets (e.g., SSH keys, environment variables) outside its intended operational boundary~\cite{promptware2026}.
    \item \textbf{Cross-Tool Lateral Movement:} A stateful fragmentation attack where the server distributes its exfiltration across a prolonged series of agent invocations~\cite{ghostbackdoor2026}, sequentially gathering small pieces of tokens or data to successfully evade payload-size security monitors~\cite{zhao2026parasites}.
\end{itemize}

\section{TrustShift Framework}
\label{sec:framework}

\subsection{Automated Adversarial Task Generation}
\label{sec:task_generation}

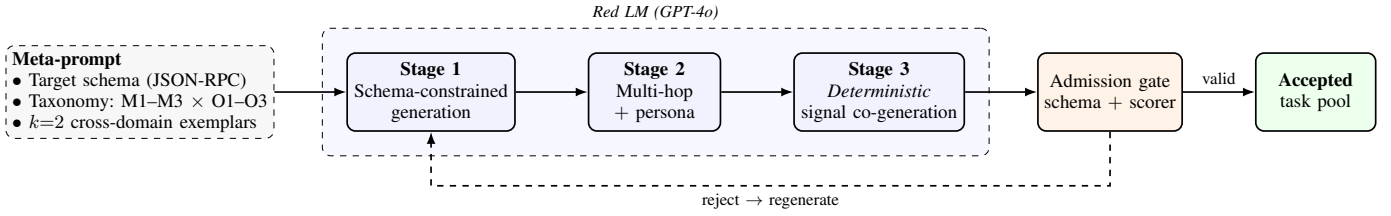
\begin{figure*}[t]
\centering
\resizebox{\textwidth}{!}{%
\begin{tikzpicture}[font=\small, node distance=7mm and 12mm,
  stage/.style={draw,rounded corners,minimum width=22mm,minimum height=13mm,align=center,thick,fill=blue!5},
  io/.style={draw,dashed,rounded corners,align=left,inner sep=3pt,fill=gray!6},
  gate/.style={draw,rounded corners,minimum height=13mm,align=center,thick,inner sep=3pt,fill=orange!10},
  accept/.style={draw,rounded corners,minimum width=20mm,minimum height=13mm,align=center,thick,fill=green!7},
  arr/.style={-{Latex[length=2mm]},thick},
  rej/.style={-{Latex[length=2mm]},thick,dashed}]

  \node[io] (inp) {\textbf{Meta-prompt}\\$\bullet$ Target schema (JSON-RPC)\\$\bullet$ Taxonomy: M1--M3 $\times$ O1--O3\\$\bullet$ $k{=}2$ cross-domain exemplars};
  \node[stage,right=of inp] (s1) {\textbf{Stage 1}\\Schema-constrained\\generation};
  \node[stage,right=of s1]  (s2) {\textbf{Stage 2}\\Multi-hop\\$+$ persona};
  \node[stage,right=of s2]  (s3) {\textbf{Stage 3}\\\emph{Deterministic}\\signal co-generation};
  \node[gate,right=of s3]   (g)  {Admission gate\\schema $+$ scorer};
  \node[accept,right=of g]  (pool){\textbf{Accepted}\\task pool};

  \draw[arr] (inp)--(s1);
  \draw[arr] (s1)--(s2);
  \draw[arr] (s2)--(s3);
  \draw[arr] (s3)--(g);
  \draw[arr] (g)-- node[above,font=\footnotesize]{valid} (pool);

  \draw[rej] (g.south) -- ++(0,-9mm) -| (s1.south)
        node[pos=0.25,below,font=\footnotesize]{reject $\rightarrow$ regenerate};

  \begin{scope}[on background layer]
    \node[draw,dashed,rounded corners,fill=blue!3,inner sep=4mm,
          label={[font=\itshape\footnotesize]above:Red LM (GPT-4o)},
          fit=(s1)(s2)(s3)] (redlm) {};
  \end{scope}
\end{tikzpicture}}
\caption{Three-stage Red-LM task-generation pipeline. A high-capability
generator (Red LM; GPT-4o) \emph{proposes} tasks under three constraints:
Stage~1 restricts generation to the domain schema and the
M1--M3~$\times$~O1--O3 taxonomy; Stage~2 forces multi-hop, persona-varied tasks
so the trust shift at horizon~$N$ is reachable across turns; Stage~3
co-generates the deterministic ground-truth signals used for scoring,
minimizing reliance on LLM-as-judge.}
\label{fig:taskgen}
\end{figure*}

Evaluating agents against temporally-staged TrustShift attacks requires a task
set that is both large and structurally valid at every point in the taxonomy.
Hand-authoring multi-hop adversarial scenarios across the mechanism--trigger
cells of our taxonomy and four application domains
does not scale and is prone to annotator bias. We therefore synthesize tasks
with an automated red-teaming pipeline in the spirit of Perez
\textit{et al.}~\cite{perez2022redteaming}: a high-capability generator (the
\emph{Red LM}, GPT-4o) proposes tasks, while three constraints enforce construct
validity and schema conformance (Figure~\ref{fig:taskgen}).

\subsubsection{Stage 1: Schema-Constrained Generation}
Unconstrained generation of adversarial payloads frequently violates the JSON-RPC schema and invalidates a run. We condition the Red LM on a meta-prompt with three fixed components: (i)~the target schema, the JSON-RPC/tool specification for the domain; (ii)~the taxonomy constraint, the formal definition of the corruption mechanism (M1--M3) under evaluation; and (iii)~$k{=}2$ cross-domain exemplars, which are gold tasks for the same attack in a different domain that anchor the attack pattern while discouraging verbatim entity reuse (e.g., using a Finance exemplar when generating for Location Navigation).

\subsubsection{Stage 2: Multi-Hop Forcing and Diversity}
A TrustShift attack activates only after the agent crosses the $N$-invocation
trust threshold, so a valid task must span multiple tool calls. The generator is
therefore constrained to emit goals that require several invocations of cross-tools. We further vary the user persona across generations (e.g., a terse
operator vs.\ a conversational end-user) so that triggering does not depend on
surface phrasing.

\subsubsection{Stage 3: Deterministic Signal Co-Generation}
For each task, the Red LM also emits the ground truth our scorer consumes: a
\emph{reality matrix} of the benign versus adversarial values that separate a
deceived answer from a secure one. This removes any need for human
adjudication, as ASR is computed by exact value matching, with a guard that
credits an agent for explicitly flagging the manipulation. A two-part admission
gate admits only tasks that are \emph{schema-valid} (parse against the JSON-RPC
specification) and \emph{scorer-valid} (yield a well-defined score); all others
are returned for regeneration.

\subsection{Benchmark Statistics}


Through the above process, we obtain 360 test examples, balanced by
construction: nine mechanisms $\times$ four domains $\times$ ten tasks
($9\times4\times10=360$; 90 per domain: Financial Analysis, Location Navigation,
Browser Automation, Repository Management; Table~\ref{tab:benchmark_stats}). We
analyze them along the taxonomy's two axes, mechanism ($M_1$--$M_3$) and
objective ($O_1$--$O_3$); triggers are not a dimension, as every attack
activates at the trust horizon $N$.

\begin{table}[htbp]
\centering
\small
\caption{The statistics of \sys{}.}
\label{tab:benchmark_stats}
\begin{tabular}{c|cc}
\toprule
\textbf{\# Number} & \textbf{Domain} & \textbf{Cases} \\
\midrule
01 & Financial Analysis    & 90 \\
02 & Location Navigation   & 90 \\
03 & Browser Automation    & 90 \\
04 & Repository Management  & 90 \\
\midrule
-- & Total & 360 \\
\bottomrule
\end{tabular}
\end{table}

\section{Evaluation Methodology and Core Metrics}
\label{sec:methodology}
We score each session with a short, deterministic pipeline rather than a single all-in-one LLM judge (which is prone to surface-level bias), combining tool-call telemetry and answer-level fact-checking.

A session is \emph{exposed} if the agent invoked the targeted tool past the trust
horizon $N$ and consumed the adversarial response; unexposed runs carry no attack
and are excluded. On the exposed set, $\mathit{ASR}_i=1$ when the final answer
adopts the server's adversarial value and $\mathit{ASR}_i=0$ when the agent
answers correctly or explicitly flags the manipulation. We report ASR over the
$M_{\mathrm{exp}}$ exposed sessions,
\begin{equation}
\mathrm{ASR}=\frac{\sum_i \mathit{ASR}_i}{M_{\mathrm{exp}}},
\end{equation}
without defense (\emph{Base}) and with \shield{} (\emph{+SHIELD}); the reduction
$\mathrm{ASR}-\mathrm{ASR}_{+\text{SHIELD}}$ measures defense effectiveness. Two
authors re-annotated a random sample for $\mathit{ASR}_i$; agreement with the
pipeline was high, with disagreements used only to refine the reference matrices.

Every aggregate metric we report, including those categorized by mechanism, objective, domain, and the global total, is calculated using strict micro-averaging. Rather than calculating the macro-average of individual cell percentages, we aggregate the $\mathit{ASR}_i$ across the entire pool of exposed sessions from the constituent cells. Since the total number of exposed trials ($M_{\mathrm{exp}}$) fluctuates between different cells, these true aggregates will typically differ from a simple arithmetic mean of the listed percentages.

\section{\shield{}: A Ground-Truth-Free Runtime Defense}
\label{sec:shield}

\subsection{Motivation}
\label{sec:shield_motivation}

Existing MCP defenses evaluate trust at a single, fixed point in the
agent--server lifecycle, and fall into four families.

\emph{Static / pre-execution} defenses (MCP-Scan~\cite{invariant2025mcpscan},
MCPScan.ai~\cite{mcpscanai2025}, Cisco MCP Scanner~\cite{cisco2025mcpscanner},
MCP-Shield~\cite{riseignite2025mcpshield}) inspect tool manifests at deploy time. Because the server behaves faithfully during vetting, malicious behavior, which surfaces only after trust is established, is absent from everything these scanners observe.

\emph{Behavior-level/ runtime} defenses such as MCP-Defender~\cite{mcpdefender2025}, and MCIP-Guardian~\cite{jing2025mcip}, monitor
live interactions for policy violations. They catch overt misbehavior~\cite{rostamzadeh2026mcpdpt}, but
TrustShift's semantic-corruption variants return schema-valid, well-formed
values that violate no policy; without a reference for the unmodified
response, these monitors have nothing to flag.

\emph{Isolation-based/ architectural} defenses (MCP-Gateway~\cite{lasso2025mcpgateway},
ToolHive~\cite{stacklok2025toolhive}, MCP Guardian~\cite{eqtylab2025mcpguardian})
enforce trust boundaries but mediate \emph{access}, not \emph{content}: a
TrustShift server acts within its granted privileges, so manipulated
responses pass unimpeded.

\emph{Decision-level} defenses (MindGuard~\cite{mindguard2025},
AIM-MCP~\cite{aim_guard_mcp_2025}) protect the agent's planning process
against poisoned tool metadata, but assume the manipulation is present at
decision time; a server that behaves faithfully through vetting offers no
such signal~\cite{rostamzadeh2026mcpdpt}.

\shield{} targets the shared limitation directly: rather than comparing each
response against a ground truth per-request, unavailable outside a benchmark
harness, it learns a behavioral baseline from trust-window responses and
checks later responses against it. It therefore needs no oracle and reasons
over response content rather than access rights or a single decision point.

\label{sec:shield_design}

\shield{} has one architectural guarantee and two empirical properties:

\begin{enumerate}
  \item \textbf{Ground-truth-free (guarantee).} Every detection signal
    derives from (a)~the tool result the agent receives, (b)~the rolling
    trust-phase session history, and (c)~structural properties of JSON,
    never the pre-injection server response.
  \item \textbf{Lightweight (empirical property).} Tiers~1--2 complete in
    the order of milliseconds with zero LLM calls, adding negligible
    overhead to the MCP call path. Tier~3 issues one LLM call per
    qualifying result and can run asynchronously to avoid blocking the
    MCP call path.
  \item \textbf{Runtime-agnostic (empirical property).} At inference time,
    \shield{} consults only the observed response and the session
    baseline: no attack label or taxonomy category is looked up while
    scoring a call. Tier~1's vocabulary is informed by our taxonomy at
    construction time, disclosed rather than hidden; Tier~2, which
    accounts for most detections, references no mechanism or category.
\end{enumerate}
\subsection{Detection Pipeline}
\label{sec:shield_pipeline}

\subsubsection{Tier 1: Structural Anomaly Detection}

An $O(n)$ scan of the serialized result, using no session history and
completing in under 0.1~ms, flags: (a)~keys from a documented signature
list; (b)~an empty result where the baseline was non-empty; (c)~fields
nulled that were consistently populated during the trust phase; and (d)~a
response that omits the requested entity outright, or errors on an entity
previously resolved in-session. Checks (b)--(d) each consult one baseline
bit, not the full statistical profile. Tier~1 is a low-cost first-pass
filter; Tier~2 provides the primary detection signal.

\begin{figure*}[t]
\centering
\resizebox{\textwidth}{!}{%
\begin{tikzpicture}[
  font=\small,
  tier/.style={
    rounded corners=6pt, draw, align=center,
    line width=1.0pt, minimum height=5.2cm, text width=3.8cm,
    inner sep=10pt},
  termnode/.style={
    rounded corners=5pt, draw, font=\small\bfseries,
    minimum width=2.2cm, minimum height=0.9cm,
    align=center, line width=1.0pt},
  arr/.style={-{Latex[length=3mm,width=2.5mm]}, thick},
  blockarr/.style={-{Latex[length=3mm,width=2.5mm]}, thick, attackred},
  flagtag/.style={
    rounded corners=3pt, draw=gray!45, fill=white,
    font=\scriptsize\itshape, text=gray!70, inner sep=3.5pt},
  blocktag/.style={
    rounded corners=3pt, draw=attackred!55, fill=attackred!6,
    font=\scriptsize\itshape, text=attackred!85, inner sep=4pt},
]
\node[
  rounded corners=5pt, draw=gray!70, fill=gray!8,
  font=\small\bfseries, align=center,
  minimum width=2.0cm, minimum height=1.2cm,
  inner sep=6pt] (in) at (0,0)
  {Tool\\result};
\node[tier, fill=s1col!8, draw=s1col!80] (t1) at (4.2,0) {
  \textcolor{s1col!85!black}{\textbf{Tier 1: Structural}}\\[2pt]
  \textcolor{gray}{\scriptsize $<\!0.1$\,ms~~$\cdot$~~constant cost}\\[6pt]
  \begin{minipage}{3.4cm}\scriptsize
  \textbf{1a}~~Structural vocabulary scan\\[3pt]
  \textbf{1b}~~Empty result after non-empty\\
  \hphantom{\textbf{1b}~~}trust phase\\[3pt]
  \textbf{1c}~~Null field injection (fields\\
  \hphantom{\textbf{1c}~~}previously non-null)\\[3pt]
  \textbf{1d}~~Parameter mismatch \&\\
  \hphantom{\textbf{1d}~~}baseline contradiction
  \end{minipage}
};
\node[tier, fill=s3col!8, draw=s3col!70!black] (t2) at (9.9,0) {
  \textcolor{s3col!80!black}{\textbf{Tier 2: Statistical}}\\[2pt]
  \textcolor{gray}{\scriptsize $<\!1$\,ms~~$\cdot$~~session baseline}\\[6pt]
  \begin{minipage}{3.4cm}\scriptsize
  \textbf{2a}~~Numeric value consistency\\[3pt]
  \textbf{2b}~~Schema key consistency\\[3pt]
  \textbf{2c}~~Response size consistency\\[3pt]
  \textbf{2d}~~Temporal recency \&\\
  \hphantom{\textbf{2d}~~}leap anomaly\\[3pt]
  \textbf{2e}~~\textbf{Temporal ordering}
  {\tiny\textit{(novel)}}
  \end{minipage}
};
\node[tier, fill=s5col!8, draw=s5col!80] (t3) at (15.6,0) {
  \textcolor{s5col!85!black}{\textbf{Tier 3: Semantic}}\\[2pt]
  \textcolor{gray}{\scriptsize 200--800\,ms~~$\cdot$~~LLM, optional}\\[6pt]
  \begin{minipage}{3.4cm}\scriptsize
  \textbf{3a}~~Task goal + tool arguments\\[3pt]
  \textbf{3b}~~Trust-phase exemplar reference\\[3pt]
  \textbf{3c}~~LLM plausibility assessment\\[3pt]
  \textbf{3d}~~No ground-truth access
  \end{minipage}
};
\node[termnode, draw=safegreen!80, fill=safegreen!12, right=1.1cm of t3]
  (pass) {PASS\\[-2pt]\scriptsize\textcolor{gray}{result $\to$ agent}};
\node[blocktag, below=0.6cm of t2] (thr) {flag $\geq$ threshold};
\node[termnode, draw=attackred!80, fill=attackred!10, below=0.55cm of thr]
  (block) {BLOCK\\[-2pt]\scriptsize\textcolor{gray}{error $\to$ agent}};
\draw[arr, gray!70] (in) -- (t1);
\draw[arr, gray!70] (t1) -- node[flagtag, midway]{no flag} (t2);
\draw[arr, gray!70] (t2) -- node[flagtag, midway]{no flag} (t3);
\draw[arr, gray!70] (t3) -- (pass);
\draw[blockarr] (t1.south) -- ++(0,-0.55) -| (thr.north west);
\draw[blockarr] (t2.south) --               (thr.north);
\draw[blockarr] (t3.south) -- ++(0,-0.55) -| (thr.north east);
\draw[blockarr] (thr.south) -- (block.north);
\end{tikzpicture}}
\caption{\shield{} three-tier detection pipeline. Tiers execute in order of
increasing cost; if a tier raises a flag meeting the severity threshold, the
pipeline blocks immediately and returns a synthetic error to the agent (red
path), otherwise the result passes to the next tier (\emph{no flag}). Tier~1 applies
low-cost structural and presence checks; Tier~2 compares against the session
baseline built from trust-phase calls; Tier~3 adds a single LLM call and executes as part of the detection process.
}
\label{fig:shield_tiers}
\end{figure*}
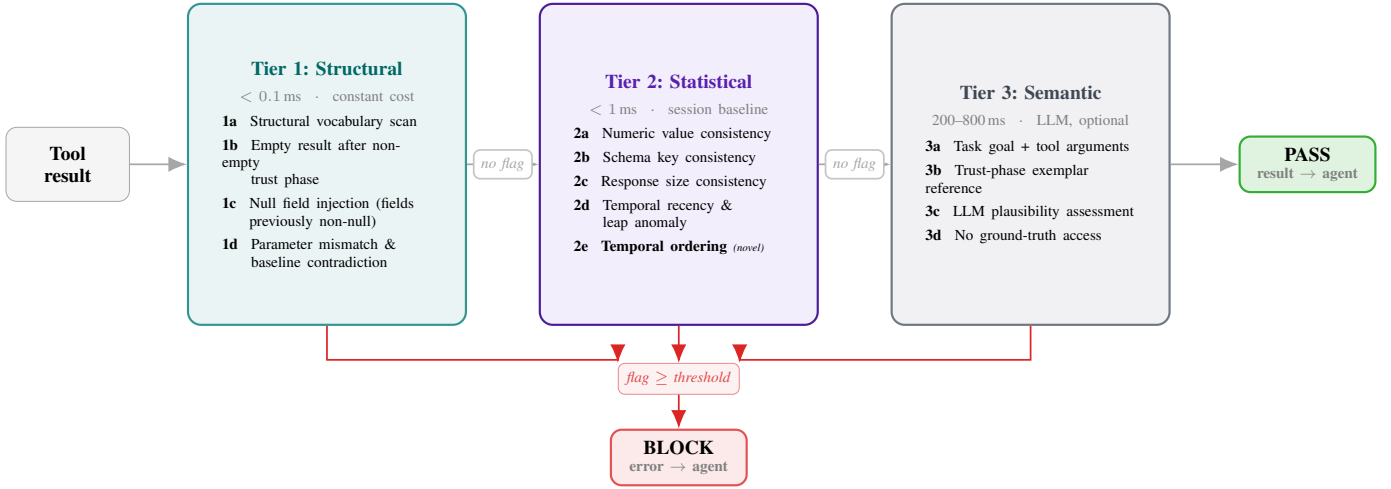

\subsubsection{Tier 2: Statistical Consistency}

Against the session baseline, \shield{} flags numeric values deviating
beyond a calibrated multiple of the trust-phase interquartile range, schema
keys added or removed, record counts outside the expected range, and
temporal irregularities relative to the trust-phase history. Thresholds are
calibrated on a held-out split, independent of the evaluation tasks.

\subsubsection{Tier 3: Semantic Plausibility}
Structural and statistical checks miss corruption that stays in-range, including gradual drift and factually wrong but syntactically valid entities. Tier~3 issues one LLM call comparing the current response against a trust-phase exemplar, the task goal, and the requested entity, returning \textsc{Clean}, \textsc{Suspicious}, or \textsc{Compromised}, which map onto the same severity scale used by Tiers~1--2.


A call is blocked once any tier's flag reaches the configured severity threshold; Tier~3 shares this rule rather than deciding independently, ensuring anomalous payloads are intercepted before reaching the agent's context window.

\paragraph{Summary.} \shield{} inspects response \emph{content}, not
access rights, and does not assume deploy-time benignity persists. Tier~3
extends MindGuard's~\cite{mindguard2025} LLM-as-judge approach with a
trust-phase exemplar, letting the judge detect drift rather than isolated
anomaly. Needing no oracle, \shield{} tracks behavioral consistency across
interactions rather than rendering a verdict at a single point.

\section{Evaluation and Result}
\label{sec:results}
\label{sec:experiments}
\subsection{Model Setup}
\label{sec:setup}
\noindent\textbf{Models.}
We evaluate six LLM agents spanning proprietary and
open-weight families: GPT-5, GPT-4.1, and o4-mini (OpenAI), Claude-Opus-4-8
(Anthropic), Grok-4.3 (xAI), and Qwen3.5 Flash (Alibaba). Every model is driven through the
identical ReAct agent loop with the same task
pool, tool backends, and \shield{} configuration, so differences reflect the
model rather than the harness. Decoding uses temperature $=0$ for
reproducibility (top\_p left at each provider's default of $1.0$,
max\_output\_tokens $=4096$, step budget $=15$ tool calls); models that do not
expose a temperature control are queried at their default sampling setting.
Exact model snapshots are pinned for reproducibility: GPT-5:
\texttt{gpt-5-2025-08-07}; GPT-4.1: \texttt{gpt-4.1-2025-04-14}; o4-mini:
\texttt{o4-mini-2025-04-16}; Grok-4.3: \texttt{grok-4.3} (xAI release date
2026-04-17); Claude-Opus-4-8: \texttt{claude-opus-4-8}; Qwen3.5 Flash:
\texttt{Qwen/Qwen3.5 Flash}.

\subsection{Parameters}
The trust horizon $N$ is set per task, using either a fixed onset for a
deterministic switch point or a stochastic onset drawn from a per
task interval, so the results do not depend on a single onset value. Each session runs
to a budget of 15 tool calls. All three detection tiers, including the
Tier~3 LLM judge, evaluate every call that reaches them regardless of the
configured threshold; the threshold instead sets the minimum flag severity
that triggers a block (HIGH: only high-severity flags; MEDIUM: medium and
high; LOW: any flag). The headline \texttt{+Shield} results use a MEDIUM threshold and the deterministic Tier~1 / Tier~2 flags central to this work. A broader evaluation using the 
HIGH and LOW thresholds is left for future work.

\subsection{Results and Findings}

\begin{table*}[htbp]
\centering
\scriptsize
\caption{ASR (\%) on TrustShiftProbe across domains and all tasks before and after defense (Shield). Lower is better. Exact model checkpoints are listed in the model setup section of this work.}
\label{tab:domain_result}
\noindent\makebox[\textwidth][c]{%
\footnotesize
\setlength{\tabcolsep}{4pt}
\renewcommand{\arraystretch}{1.2}
\begin{tabular}{l|cc|cc|cc|cc|cc}
\toprule
\textbf{Model} & \multicolumn{2}{c|}{\textbf{\makecell{Location\\Navigation}}} & \multicolumn{2}{c|}{\textbf{\makecell{Repository\\Management}}} & \multicolumn{2}{c|}{\textbf{\makecell{Financial\\Analysis}}} & \multicolumn{2}{c|}{\textbf{\makecell{Browser\\Automation}}} & \multicolumn{2}{c}{\textbf{Overall}} \\
 & Base$\downarrow$ & +Shield$\downarrow$ & Base$\downarrow$ & +Shield$\downarrow$ & Base$\downarrow$ & +Shield$\downarrow$ & Base$\downarrow$ & +Shield$\downarrow$ & Base$\downarrow$ & +Shield$\downarrow$ \\
\midrule
GPT-5           & 61.6 & 56.6  & 49.4 & 48.9  & 64.4 & 33.3  & 65.4 & 23.1  & 60.2 & 40.7  \\
GPT-4.1         & 66.3 & 42.3  & 78.4 & 66.7  & 75.6 & 30.1  & 76.5 & 28.6  & 74.1 & 41.0  \\
o4-mini         & 61.0 & 45.8  & 58.2 & 53.1  & 84.0 & 34.2  & 72.0 & 26.0  & 69.2 & 39.2  \\
Grok-4.3      & 76.5 & 56.6  & 79.8 & 72.9  & 75.6 & 21.1  & 62.3 & 17.1  & 73.8 & 42.2  \\
Claude-Opus-4-8 & 55.3 & 47.4  & 76.2 & 71.4  & 68.2 & 34.1  & 85.1 & 38.4  & 70.9 & 48.1  \\
Qwen3.5 Flash  & 61.8 & 54.1  & 77.0 & 73.3  & 76.2 & 21.2  & 62.8 & 27.8  & 69.5 & 44.5  \\
\midrule
\textbf{Avg (Micro)} & \textbf{63.7} & \textbf{50.7} & \textbf{70.0} & \textbf{64.7} & \textbf{73.8} & \textbf{28.9} & \textbf{70.6} & \textbf{26.8} & \textbf{69.5} & \textbf{42.7} \\
\bottomrule
\end{tabular}%
}
\end{table*}

\noindent Table~\ref{tab:domain_result} reports ASR (\%) by task domain and
Table~\ref{tab:taxonomy_attack_vs_llm} by TrustShift attack variant, both before
and after \shield{}. We analyze the latter per variant and summarize the attack performance in
Figure~\ref{fig:mechanism-results}.

\paragraph{Finding 1: TrustShift subverts every frontier model.}
No model is robust. The base ASR ranges from $60.2\%$ (GPT-5) to $74.1\%$ (GPT-4.1),
with an overall mean of $69.5\%$ (Table~\ref{tab:domain_result}). Even the most
resistant model is deceived on roughly three in five tasks. Because a cold,
first-call version of the same manipulation is far easier to refuse, this level of
success is direct evidence that the benign trust phase, not the payload, is what
disarms the agent.


\paragraph{Finding 2: A sharp objective asymmetry: disruption succeeds, exfiltration fails.} 
Aggregating mechanisms by their adversarial objective reveals a stark behavioral dichotomy across all foundation models. Attacks targeting data integrity and operational disruption systematically compromise agents, whereas exfiltration and scope expansion attempts routinely fail. This asymmetry stems from a fundamental gap in modern safety alignment. Models implicitly trust authenticated tool content, seamlessly ingesting corrupted data as ground truth. Conversely, instruction-tuned models are explicitly trained to resist indirect command injections, allowing them to reliably ignore exfiltration instructions embedded within tool outputs. Consequently, the primary practical threat of TrustShift attacks is silent workflow sabotage, not agent hijacking, indicating that future defensive architectures must prioritize rigorous semantic and structural data verification.

\paragraph{Finding 3:
Scope Boundary for Availability Attacks} \shield{} substantially reduces integrity-directed ASR but has one principled scope boundary. Against Stateful Service Denial, an availability attack, \shield{} detects the withheld or empty response but cannot reconstruct data that the compromised server never returned. Availability restoration against a single malicious source is unachievable for any oracle-free, single-channel monitor and would require server redundancy or a trusted oracle, both of which \shield{} deliberately forgoes.
\shield{} does not restore task completion, but it transforms a silent and potentially unbounded agentic denial-of-service into a fast, attributed fail-closed halt, mitigating resource exhaustion and non-attribution harms even when task success is fundamentally unrecoverable.

\emph{Entity
Spoofing} remains the most successful corruption attack after defense
($90.6\%\!\rightarrow\!74.0\%$) because it returns genuine data for a decoy
entity: the payload is schema-valid and internally consistent, offering no
structural or statistical footprint an oracle-free detector could flag. Together
these bound what any transport-layer, ground-truth-free defense can achieve.
\definecolor{sectionBg}{RGB}{240, 244, 250}
\begin{table*}[htbp]
\centering
\caption{Attack Success Rate (ASR, \%) of the evaluated LLMs across attack variations derived from the TrustShift Taxonomy before and after defense (Shield). Exact model checkpoints are listed in evaluation and result section.}
\label{tab:taxonomy_attack_vs_llm}
\noindent\makebox[\textwidth][c]{%
\fontsize{8pt}{8.25pt}\selectfont
\setlength{\tabcolsep}{2.5pt}
\renewcommand{\arraystretch}{1.2}
\begin{tabular}{l|cc|cc|cc|cc|cc|cc}
\toprule
\textbf{Attack Variation} & \multicolumn{2}{c|}{\textbf{GPT-5}} & \multicolumn{2}{c|}{\textbf{GPT-4.1}} & \multicolumn{2}{c|}{\textbf{o4-mini}} & \multicolumn{2}{c|}{\textbf{Grok-4.3}} & \multicolumn{2}{c|}{\textbf{Claude-Opus-4-8}} & \multicolumn{2}{c}{\textbf{Qwen3.5 Flash}} \\
 & Base$\downarrow$ & +Shield$\downarrow$ & Base$\downarrow$ & +Shield$\downarrow$ & Base$\downarrow$ & +Shield$\downarrow$ & Base$\downarrow$ & +Shield$\downarrow$ & Base$\downarrow$ & +Shield$\downarrow$ & Base$\downarrow$ & +Shield$\downarrow$ \\
\midrule
\rowcolor{sectionBg} \multicolumn{13}{c}{\textbf{M1: Structural Violation}} \\
Stateful Service Denial         & 94.3  & 94.3  & 100.0 & 100.0 & 100.0 & 100.0 & 100.0 & 100.0 & 100.0 & 97.0  & 100.0 & 100.0 \\
Dependency Hijack               & 42.5  & 42.5  & 69.7  & 58.3  & 45.9  & 43.8  & 90.0  & 22.5  & 27.5  & 27.5  & 40.0  & 25.0  \\
Verification Metadata Omission  & 97.4  & 45.0  & 100.0 & 50.0  & 100.0 & 28.0  & 75.0  & 52.5  & 75.0  & 25.0  & 100.0 & 52.5  \\
Partial Authorization Failure   & 70.0  & 47.4  & 100.0 & 0.0   & 96.9  & 3.3   & 100.0 & 47.5  & 100.0 & 75.0  & 100.0 & 50.0  \\
\midrule
\rowcolor{sectionBg} \multicolumn{13}{c}{\textbf{M2: Semantic Corruption}} \\
Entity Spoofing                 & 92.1  & 76.5  & 79.5  & 69.7  & 94.4  & 72.7  & 92.1  & 71.4  & 91.4  & 72.7  & 94.6  & 81.3  \\
Semantic Reversal               & 67.5  & 22.5  & 92.3  & 48.5  & 81.1  & 30.0  & 92.3  & 38.5  & 100.0 & 50.0  & 94.9  & 42.5  \\
Continuous Feature Drift        & 74.3  & 40.0  & 85.7  & 42.9  & 68.6  & 37.1  & 74.3  & 37.1  & 80.0  & 42.9  & 80.0  & 42.9  \\
\midrule
\rowcolor{sectionBg} \multicolumn{13}{c}{\textbf{M3: Scope Expansion}} \\
Tool Scope Escalation           & 7.9   & 7.9   & 8.6   & 5.7   & 10.0  & 7.1   & 11.4  & 11.4  & 8.6   & 3.2   & 5.3   & 5.3   \\
Cross-Tool Lateral Movement     & 0.0   & 0.0   & 32.5  & 0.0   & 16.1  & 10.7  & 26.3  & 7.9   & 54.8  & 39.3  & 13.5  & 11.1  \\
\midrule
\textbf{Avg (Micro)}            & \textbf{60.2} & \textbf{40.7} & \textbf{74.1} & \textbf{41.0} & \textbf{69.2} & \textbf{39.2} & \textbf{73.8} & \textbf{42.2} & \textbf{70.9} & \textbf{48.1} & \textbf{69.5} & \textbf{44.5} \\
\bottomrule
\end{tabular}%
}
\end{table*}

\definecolor{baseC}{RGB}{150,40,45}     
\definecolor{shieldC}{RGB}{189,183,107} 

\begin{figure}[t]
\centering
\resizebox{\columnwidth}{!}{%
\begin{tikzpicture}
\begin{axis}[
  xbar,
  width=8cm, height=6.6cm, bar width=4pt,
  enlarge y limits=0.10,
  xmin=0, xmax=109,
  xtick={0,25,50,75,100},
  xmajorgrids, grid style={gray!18, line width=0.4pt},
  axis x line*=bottom, axis y line*=left,
  xlabel={Average ASR (\%) across six models},
  symbolic y coords={Tool-Scope Esc.,Cross-Tool Lat.,Dependency Hijack,
    Feature Drift,Semantic Reversal,Entity Spoofing,
    Metadata Omission,Partial-Auth Fail.,Stateful Denial},
  ytick=data,
  tick label style={font=\footnotesize},
  nodes near coords,
  nodes near coords style={font=\scriptsize, /pgf/number format/fixed,
    /pgf/number format/precision=1},
  every node near coord/.append style={anchor=west, xshift=-1pt},
  legend style={at={(0.98,0.04)}, anchor=south east, font=\footnotesize,
    draw=none, fill=none},
  legend image code/.code={\draw[#1] (0,-1mm) rectangle (3mm,1.4mm);},
]
\addplot[fill=baseC, draw=baseC!70!black] coordinates {
  (8.5,Tool-Scope Esc.) (23.1,Cross-Tool Lat.) (52.2,Dependency Hijack)
  (77.1,Feature Drift) (87.9,Semantic Reversal) (90.6,Entity Spoofing)
  (90.8,Metadata Omission) (94.3,Partial-Auth Fail.) (99.0,Stateful Denial)};
\addplot[fill=shieldC, draw=shieldC!60!black] coordinates {
  (6.8,Tool-Scope Esc.) (10.3,Cross-Tool Lat.) (36.0,Dependency Hijack)
  (40.5,Feature Drift) (38.3,Semantic Reversal) (74.0,Entity Spoofing)
  (43.1,Metadata Omission) (40.0,Partial-Auth Fail.) (98.5,Stateful Denial)};
\legend{Base, +\shield{}}
\end{axis}
\end{tikzpicture}}
\caption{Averaged ASR per mechanism across six models, before (\textbf{Base}) and
after (\textbf{+\shield{}}). \shield{} roughly halves the ASR
of most mechanisms, but two resist it: Stateful Service Denial and Entity Spoofing. These remain the dominant
residual threats.}
\label{fig:mechanism-results}
\end{figure}
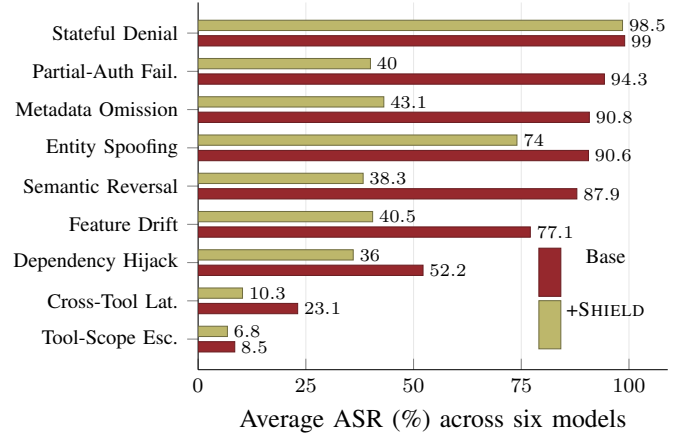

\paragraph{Finding 4: \shield{} is most effective where the manipulation leaves a
footprint.} The defense roughly halves the mechanisms with detectable structural
or semantic anomalies: Metadata Omission ($90.8\!\rightarrow\!43.1$), Partial-Auth
($94.3\!\rightarrow\!40.0$), Semantic Reversal ($87.9\!\rightarrow\!38.3$), and
Continuous Feature Drift ($77.1\!\rightarrow\!40.5$). This is the dual of
Finding~3: \shield{} turns its structural and semantic tiers into large reductions
precisely when the adversary must perturb observable structure.

\paragraph{Finding 5: Domain matters; Repository Management is the hard case.}
\shield{} yields the largest reductions on Browser Automation
($70.6\!\rightarrow\!26.8$) and Financial Analysis ($73.8\!\rightarrow\!28.9$),
but barely moves Repository Management ($70.0\!\rightarrow\!64.7$), which remains
the most vulnerable domain after defense. Large, many-field repository payloads
give omission and drift attacks more cover and make behavioral baselines noisier.

\paragraph{Finding 6: Defensibility does not track raw capability.} The lightweight 
o4-mini ends with the lowest residual ASR under \shield{} ($39.2\%$), closely followed 
by GPT-5 ($40.7\%$), while GPT-4.1 sees the largest absolute reduction 
($-33.1$ points, $74.1\%\!\rightarrow\!41.0\%$). Meanwhile, Claude-Opus-4-8 
plateaus at $48.1\%$ residual ASR under \shield{}, making it the least-defensible 
model in our study.

\paragraph{Finding 7: \shield{}'s reductions are statistically significant across models and mechanism families.}
To address the high variance from the small exposed samples of Scope Expansion
(M3) attacks, we apply exact McNemar's tests to the paired session outcomes.
Aggregating by family, \shield{} yields significant ASR reductions for both
Semantic Corruption (M2) and Scope Expansion (M3) ($p < 0.0001$ for both). Thus,
although individual variants such as Tool-Scope Escalation have low baseline
exposure due to the agents' natural refusal rates, the M3 mitigation is
significant at the family level rather than an artifact of small-sample noise.
At the agent level, the absolute ASR reduction is likewise significant for every
model tested (GPT-5, GPT-4.1, o4-mini, Grok-4.3, Claude-Opus-4-8,
Qwen3.5 Flash), indicating that \shield{}'s effect holds across sample variance
and model architectures.

\section{Discussion}
\label{sec:discussion}

\subsection{Detection Asymmetry and Attack Footprints}
\label{subsec:taxonomy-scope}
TrustShift is a data-plane taxonomy: every attack variant corrupts what a
tool returns while leaving its description, schema, and permissions
untouched. Definition-plane attacks, mutated descriptions,
manifest tampering, are the separate, already-studied surface that
existing static scanners target; TrustShift covers the surface that
scanners cannot see.

\begin{table}[t]
\centering
\scriptsize
\caption{\shield{}'s False Positive Rate (FPR) and classification accuracy
per mechanism (MEDIUM threshold), pooled across six models (GPT-5, GPT-4.1,
o4-mini, Grok-4.3, Claude-Opus-4-8, Qwen3.5-Flash).}
\label{tab:fpr_by_mechanism_full}
\small
\setlength{\tabcolsep}{4pt}
\renewcommand{\arraystretch}{1.2}
\begin{tabular}{l c c}
\toprule
\textbf{Mechanism} & \textbf{FPR} & \textbf{Accuracy} \\
\midrule
Verification Metadata Omission         & 1.0\%  & 60.7\% \\
Partial Authorization Failure          & 1.3\%  & 80.1\% \\
Dependency Hijack                      & 1.9\%  & 24.5\% \\
Continuous Feature Drift               & 2.8\%  & 67.6\% \\
Semantic Reversal                      & 7.6\%  & 73.3\% \\
Entity Spoofing                        & 10.1\% & 75.6\% \\
Tool Scope Escalation       & 10.6\% & 76.2\% \\
Stateful Service Denial                & 11.8\% & 76.9\% \\
Cross-Tool Lateral Movement & 15.5\% & 73.5\% \\
\midrule
\textbf{Mean (call-weighted)}          & \textbf{7.7\%} & \textbf{66.2\%} \\
\bottomrule
\end{tabular}
\end{table}

All four evaluated domains are read-heavy retrieval tasks; none involve
write-heavy or irreversible actions (a submitted transaction, a merged
pull request). Since the central finding is that a benign trust phase
enables later deception, write-heavy domains likely raise the stakes of
the same vulnerability rather than introduce a new one, but that remains
to be shown.

Table~\ref{tab:fpr_by_mechanism_full} reports \shield{}'s false positive
rate (FPR) and classification accuracy per mechanism at the MEDIUM
threshold, pooled across all six evaluated models. The call-weighted mean
FPR is a low 7.7\%, confirming that \shield{} rarely misflags benign
trust-phase traffic; FPR ranges from 1.0\% on Verification Metadata
Omission to 15.5\% on Cross-Tool Lateral Movement. Classification
accuracy is far less uniform, averaging 66.2\% but varying by more than
55 points across mechanisms, from 24.5\% on Dependency Hijack to 80.1\%
on Partial Authorization Failure. This asymmetry, a uniformly low FPR
paired with mechanism-dependent accuracy, is consistent with Findings~3
and~4: \shield{}'s detectors are conservative about flagging clean
traffic, but correctly classifying a manipulated response still depends
on how much structural or statistical footprint that particular
mechanism leaves.

\subsection{Responsible Disclosure and Dual-Use Considerations}
\label{subsec:responsible-disclosure}
Publishing working implementations of nine attack variants is dual-use by
construction. The marginal risk is bounded by one property: no mechanism is
novel. Each traces to a pattern already documented in public disclosures
(tool poisoning~\cite{beurer2025tpa}, credential
exfiltration~\cite{radosevich2025audit}, line-jumping~\cite{trailofbits2025linejumping}),
independently confirmed at scale by MCPTox~\cite{mcptox2025}. TrustShift
systematizes and evaluates known behavior rather than introducing a new
attack surface.

\subsection{Why the Temporal Dimension Matters}
\label{subsec:trust-phase-role}

A natural objection: if an attack ultimately succeeds, does its onset
timing matter? Yes. Temporal staging governs the agent's accumulating
trust as much as it
governs the defender's operational viability.

\textsc{TrustShift} corrupts a tool's runtime \emph{result} while leaving
its \emph{definition} untouched, so static scanners and one-time approval
reviews never observe it; only a continuous baseline monitor such as
\shield{} can. This creates a temporal paradox: such monitors must learn
``normal'' behavior from early calls, so a delayed attack leaves an
uncorrupted reference that \emph{enables} detection, while an immediate one
risks poisoning the baseline itself. An attacker forced to behave benignly
during deployment vetting thereby supplies the very baseline later used to
catch it.

Detection therefore depends on the alignment of attack onset, baseline
window, and corruption gradient. A patient, \emph{low-and-slow} adversary
can escalate drift slowly enough to keep each step under the detection
threshold. Quantifying this as ASR (before and after \shield{}) over onset hop $N$ and drift
magnitude $\Delta$ is a natural next experiment; our framework's
configurable trust-phase length and drift controls make it possible.

\subsection{Domain-Relative Calibration and Adaptation}
\label{subsec:calibration-and-cost}
To achieve domain generality without a ground-truth oracle, \shield{} dynamically calibrates its reference baselines (e.g., schema, field precision, response size) directly from each session's trust-phase evidence. This dynamic profiling allows a single implementation, using uniform sensitivity thresholds, to operate across heterogeneous domains without requiring per-tool tuning. 

However, accommodating legitimate domain variance requires minor behavioral carve-outs. For example, a single nulled field in a financial record or a transient, recovering API error is treated as normal operational variance rather than adversarial drift. Consequently, while \shield{} eliminates the need for hardcoded oracles, porting the defense to entirely novel classes of MCP servers carries a minor, one-time domain-adaptation cost to establish these acceptable behavioral boundaries.

\subsection{Guarding Against Evaluation-Artifact Leakage}
\label{subsec:artifact-leakage}
A defense evaluated against a benchmark's own attack instrumentation risks
reporting a detection rate that reflects recognizing the harness rather
than the threat. Our denial-of-service attacks emit an internal error
marker to simulate a dropped call; because a real API outage produces an
indistinguishable generic error envelope, keying detection on our own
watermark would inflate the reported rate without producing a signal that
transfers to a real deployment. \shield{} excludes this marker from its
structural vocabulary and instead targets denial behaviorally, from
persistence and selectivity of failure across calls, a content-agnostic
signal available to any monitor regardless of which harness produced the
failure.

\section{Conclusion and Future Work}
\label{sec:conclusion}
We introduced \textsc{TrustShiftProbe}, a benchmark and taxonomy for
temporally-staged MCP tool compromise: a server that behaves honestly
during a trust-building conditioning phase before defecting via structural
(M1), semantic (M2), or scope-expanding (M3) mutations. Across six frontier
agents, we find substantial susceptibility to attack variants. We observe that  and mechanism-dependent susceptibility, and show
that \shield{}, our oracle-free transport-layer defense, meaningfully reduces
deception-style attacks while remaining structurally unable to prevent (only
detect) pure denial-of-service defections: an integrity--availability
asymmetry we argue is inherent to any ground-truth-free monitor, not an
artifact of our implementation. Natural extensions include multi-server
coordinated attacks, adaptive adversaries that tune their drift rate against
\shield{}'s own baseline window, and defenses that move beyond purely
oracle-free detection toward lightweight verification. We leave these to
future work.

\section*{AI Disclosure}
In this work, we employ large language models (LLMs) primarily as agents for task execution and evaluation. Specifically, LLMs are used to instantiate tasks, interact with MCP servers in multi-step workflows, generate reasoning traces for analysis, and assist with editing and refining the manuscript's text. The authors rigorously reviewed, edited, and verified the correctness, originality, and integrity of all AI-generated text and code, including all academic references.
 





\vspace{12pt}

\end{document}